# Debunking Seven Myths about 5G New Radio

Xingqin Lin

Ericsson Research, Santa Clara, CA, USA

Contact: xingqin.lin@ericsson.com

**ABSTRACT**

New radio (NR) is a new wireless access technology developed as part of the fifth-generation (5G) of mobile communications to support a wide range of services, devices, and deployments. NR features spectrum flexibility, ultra-lean design, forward compatibility, low latency support, and advanced antenna technologies. There has been excitement about NR, sometimes clouded by confusion. This article is an attempt to summarize and overview the key features of NR by debunking seven of the more popular myths and revealing what NR really is. The seven topics include spectrum, flexible waveform and multiple access, LTE-NR interworking and coexistence, low latency support, massive machine type communications, non-terrestrial communications, and beyond radio.

**INTRODUCTION**

5G is the fifth-generation of mobile communications, addressing a wide range of use cases from enhanced mobile broadband (eMBB) to ultra-reliable low-latency communications (URLLC) to massive machine type communications (mMTC). The third-generation partnership project (3GPP), a global standard-development organization for mobile communications, has developed a new air interface – new radio (NR) – for 5G networks [1]. There has been excitement about NR, sometimes clouded by confusion. In this article, we identify seven common myths, debunk them, and reveal what NR really is.

In early 2012, the working party 5D (WP 5D) in the International Telecommunication Union (ITU) Radiocommunication Sector (ITU-R) initiated a program to develop the International Mobile Telecommunications (IMT) for 2020 and beyond (IMT-2020), whose framework and objectives are outlined in the ITU-R Recommendation M.2083 [2]. This sparked much interest and set the stage for 5G research activities worldwide [3].

Meanwhile, there was emerging consensus in the mobile industry that a new radio access technology would be needed as part of 5G. The momentum led to a 3GPP 5G workshop held in September 2015, in conjunction with the 3GPP radio access network (RAN) plenary meeting #69 in Phoenix, AZ, USA. This workshop marked the beginning of the NR work in 3GPP. It was emphasized at this workshop that the design of NR should be forward compatible to facilitate the introduction of new technologies and applications in the future [4].

After the 5G workshop, 3GPP conducted several studies in Release 14 and finished the first release of NR specifications as part of Release 15. The Release-15 NR specifications can fulfill a subset of the IMT-2020 requirements. 3GPP continues NR evolution in Release 16, which will fulfill all the IMT-2020 requirements.

NR features spectrum flexibility, ultra-lean design, forward compatibility, low latency support, and advanced antenna technologies. A detailed introduction to NR can be found in [5]. This article summarizes seven key features of NR and discusses how they have been somewhat misinterpreted.

**Spectrum.** A distinct feature of NR is the support of operation at millimeter wave frequencies to reap the benefits of large amounts of spectrum for providing extreme data rates. However, NR operation at low-band (below 1 GHz) and mid-band (1 – 7 GHz) is important as well.

**Flexible waveform and multiple access.** NR adopts cyclic-prefix orthogonal frequency division multiplexing (CP-OFDM) for both downlink and uplink transmissions, and additionally supports the use of discrete Fourier transform (DFT) spread OFDM (DFT-S-OFDM) in the uplink. Accordingly, orthogonal frequency division multiple access (OFDMA) is the natural multiple access scheme in NR. NR features flexible OFDM numerology to support a wide range of deployment scenarios and carrier frequencies.

**LTE-NR interworking and coexistence.** LTE-NR dual connectivity, in which devices may have simultaneous LTE and NR connectivity, enables tight LTE-NR interworking. NR and LTE can be deployed in the same spectrum with dynamic spectrum sharing, facilitating a smooth 5G rollout.

**Low latency support.** An important design consideration of NR is latency reduction. NR encompasses a rich set of tools such as mini-slot transmission, tightened device processing

| Aspect | Myth | Fact |
|---|---|---|
| **Spectrum** | NR = Millimeter wave. | NR supports operation in a wide range of spectrum, from sub-6 GHz to millimeter wave frequencies. Combining the uses of low-band, mid-band, and high-band spectrum in NR deployments achieves highest quality performance. |
| **Flexible waveform and multiple access** | New non-CP-OFDM waveform and non-orthogonal multiple access are foundations of NR. | NR adopts CP-OFDM for both downlink and uplink transmissions, and additionally supports DFT-S-OFDM in the uplink. OFDMA is the primary multiple access scheme in NR. |
| **LTE-NR interworking and coexistence** | With NR, LTE is not needed. | LTE and NR will coexist in the foreseeable future. Tight LTE-NR interworking and dynamic spectrum sharing between LTE and NR are key to the smooth migration towards 5G. |
| **Low latency support** | 5G delivers 1 ms end-to-end latency everywhere. | IMT-2020 over-the-air latency requirements of user plane are 1 ms and 4 ms for URLLC and eMBB, respectively. NR encompasses a rich set of tools to achieve low latency. |
| **Massive machine type communications** | NR is the only 5G access technology to serve massive machine type communications. | While NR can serve many mMTC use case, NB-IoT and LTE-M are the more optimized 5G radio access technologies for LPWA connectivity. NR can well coexist with NB-IoT and LTE-M. |
| **Non-terrestrial communications** | NR is all about terrestrial mobile communications. | Enabling NR to support non-terrestrial communications with satellites and high-altitude platform stations is a direction under exploration. Terrestrial NR networks can provide connectivity to UAVs in the sky besides serving terrestrial UEs. |
| **Beyond radio** | 5G = NR. | 5G goes beyond radio. The entire system of 5G is being transformed towards enabling new services and new ecosystems. |

*Table 1: Summary of the debunked seven myths and key features about NR.*

times, and optimized higher-layer protocols, to enable low latency support.

**Massive machine type communications.** NR can serve many mMTC use case, but narrowband internet of things (NB-IoT) and long-term evolution for machines (LTE-M) are the more optimized 5G radio access technologies for low power wide area (LPWA) connectivity. Further, NB-IoT and LTE-M can be deployed in NR frequency band, allowing for harmonious coexistence of NB-IoT/LTE-M with NR.

**Non-terrestrial communications.** To further expand the 5G ecosystem, 3GPP has been studying the possibility of evolving NR to support non-terrestrial communications (NTN) with satellites and high-altitude platform stations. In addition, NR can be used to provide connectivity to unmanned aerial vehicles (UAVs).

**Beyond radio.** While radio is a prominent component of 5G, transformations in the other parts of 5G system including core network, backhaul, and fronthaul are also key to enabling new services, new ecosystems, and new revenues in the 5G era.

With the help of advanced features (including the aforementioned ones), NR allows for deployments in a wide range of scenarios for supporting different use cases. Commercial NR networks are starting to go live across the world. Large scale commercial deployments are happening in North America, Asia, and Europe. However, several key features of NR have been somewhat misinterpreted, into what we call the seven myths. The rest of this article debunks the seven common myths and discusses the key NR design features in more detail, with a summary given in Table 1.

## SPECTRUM

**Myth 1:** *NR = Millimeter wave.*

There has been much excitement about using millimeter wave spectrum for 5G cellular [6], due to the large amounts of spectrum at millimeter wave frequencies that can offer large system capacity and high data rates. Millimeter wave has drawn much attention (and rightfully so) such that sometimes 5G operations in low-band and mid-band spectrum are forgotten, resulting in a misconception that NR is all about millimeter wave communication.

Low-band spectrum below 1 GHz is ideal for providing wide-area and deep indoor coverage. The newly allocated low-band spectrum for IMT includes the 600 MHz band and the 700 MHz band. Mid-band spectrum ranging from 1 GHz to 7 GHz can provide carrier bandwidths up to 100 MHz. It strikes a good balance between coverage, capacity, data rate, and latency. The 3.5 GHz band is an exemplary new global band for IMT.

NR features spectrum flexibility and supports operation in the spectrum ranging from sub-1 GHz to millimeter wave bands. Two frequency ranges (FR) are defined in Release 15: FR1: 410 MHz – 7.125 GHz; FR2: 24.25 GHz – 52.6 GHz. While the use of mid-band and high-band spectrum allows for larger channel bandwidths, operation at higher frequencies needs to tackle the increased propagation losses.

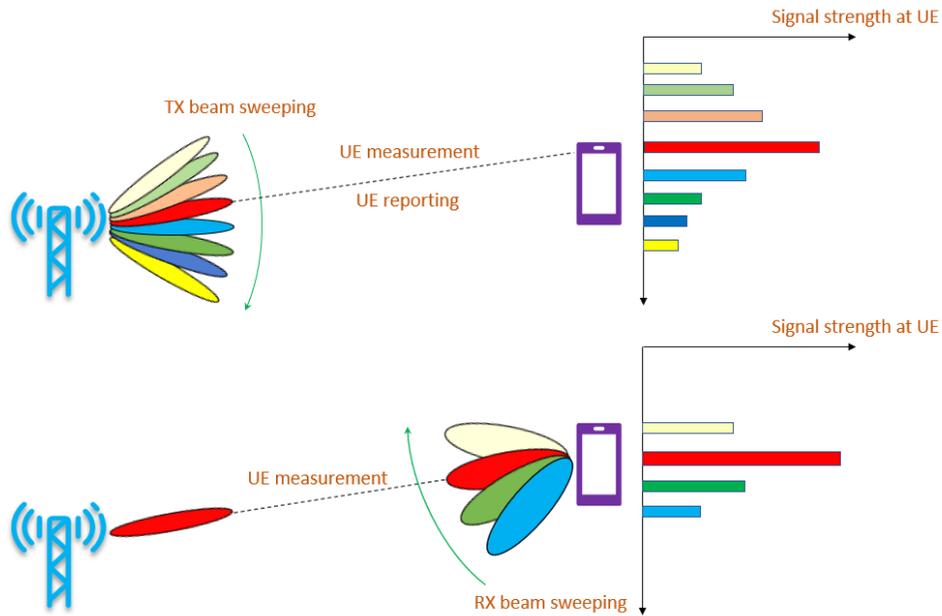

*Figure 1: An illustration of downlink transmitter side and receiver side beam adjustment in NR.*

Propagation losses at higher frequencies are not insurmountable. NR adopts a beam centric design and exploits high-gain antennas at both transmitters and receivers to beamform signal energy. The NR channels, signals, and procedures have been designed to support beamforming. Several key beam centric design aspects are in order.

- *Cell search:* NR defines synchronization signal block (SSB), consisting of the primary synchronization signal (PSS), the secondary synchronization signal (SSS), and the physical broadcast channel (PBCH). It is possible to apply beam sweeping to transmit SSBs in different beams at different times. Within an SSB burst set period, up to 64 SSBs can be transmitted in different beams.
- *Random access:* Different random-access occasions and/or different preambles can be configured to be associated with different SSBs. User equipment (UE) can choose the right random-access preamble to transmit in the uplink based on the SSB it acquires in the downlink. A 5G NodeB (gNB) can determine the beam direction based on the received preamble. This allows for establishing a suitable beam pair during the initial access procedure.
- *Beam tracking:* The beam pair established during the initial access procedure needs to be regularly reassessed and adjusted to cope with movements of UE and other objects in the propagation environment. The downlink transmitter side beam adjustment is based on UE measurements and reporting on a set of reference signals (channel-state information reference signal (CSI-RS) or SSB). The downlink receiver side beam adjustment is also carried out based on the UE measurements of downlink reference signals, but UE reporting of the measurement results is not needed. Figure 1 gives an illustration of the downlink beam adjustment scheme in NR. The uplink beam adjustment can be implicit assuming downlink/uplink beam correspondence or explicit based on configured sounding reference signal (SRS) in the uplink.
- *Beam indication:* NR introduces a so-called transmission configuration indication (TCI) mechanism for beam indication. Each TCI state includes information about a downlink reference signal (CSI-RS or SSB). The gNB can indicate a TCI state to inform the UE that the physical downlink control channel (PDCCH) or the physical downlink shared channel (PDSCH) is transmitted using the same beam as the configured reference signal.
- *Beam recovery:* NR includes beam-failure recovery functionality to allow for reestablishment of a suitable beam pair upon a beam failure event. The procedure includes UE detecting the beam failure event, identifying a new candidate beam pair, requesting for beam recovery, and gNB response to the request.

In a nutshell, NR features a beam centric design to support operation in a large range of spectrum. Each spectrum band features unique properties. It is essential to combine the uses of low-band, mid-band, and high-band spectrum in NR deployments to achieve highest quality performance.

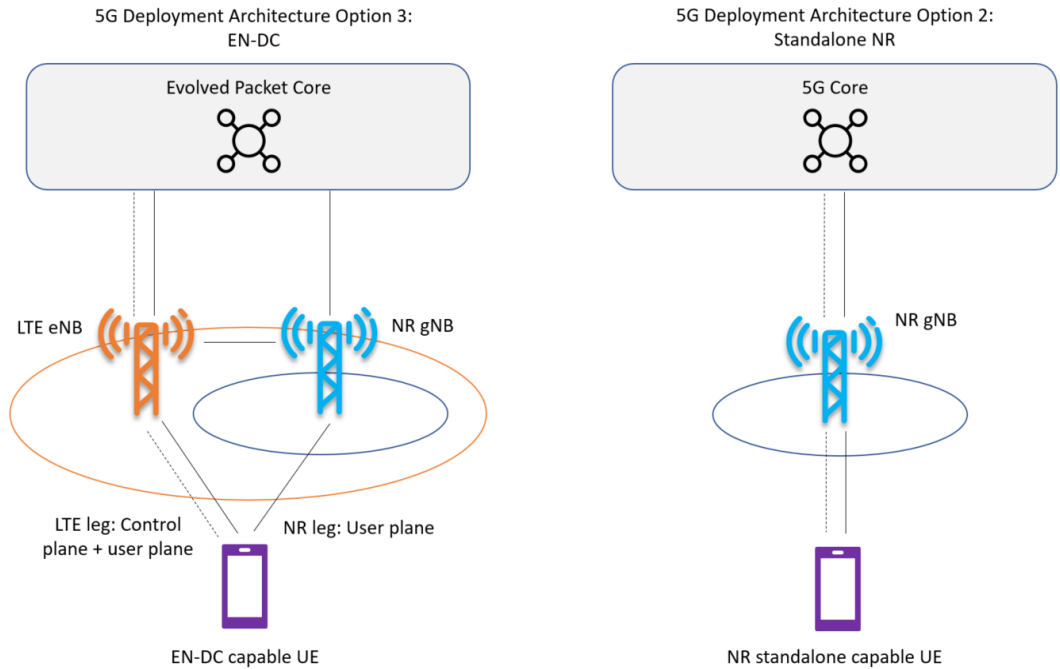
Figure 2: An illustration of 5G deployment Option 3 (EN-DC) and Option 2 (standalone NR).

## FLEXIBLE WAVEFORM AND MULTIPLE ACCESS

**Myth 2:** *New non-CP-OFDM waveform and non-orthogonal multiple access are foundations of NR.*

Waveform is a fundamental aspect of any wireless communications system. On the road to NR, there were abundant "non-CP-OFDM" waveform proposals, such as filter-bank multicarrier (FBMC), generalized frequency-division multiplexing (GFDM), and different variations of CP-OFDM by applying additional filtering, windowing, or precoding. Much work showed that CP-OFDM was outperformed in one way or the other by the new waveform proposal, leading to a myth that NR would use a non-CP-OFDM based waveform.

The new waveform proposals were often motivated by overcoming the disadvantages of OFDM in peak-to-average-power ratio (PAPR) and out-of-band emission to reduce power backoff, lower spectral regrowth, and relax synchronization requirements. The benefits however come at the cost of much increased transceiver complexity and incompatibility with multiple-input multiple-output (MIMO) techniques.

As a mature technology, CP-OFDM enjoys important advantages including high spectral efficiency, compatibility with MIMO, and low implementation complexity. The high PAPR and lower frequency localization can be mitigated to some extent by well-established techniques, such as clipping, filtering, and windowing. As a result, after assessments of all the waveform proposals, 3GPP adopted CP-OFDM for both downlink and uplink transmissions in NR. NR additionally supports DFT-S-OFDM in the uplink to improve coverage.

To support NR deployment in a wide range of deployment scenarios and carrier frequencies, NR defines scalable OFDM numerologies using subcarrier spacing of $2^{\mu} \cdot 15$ kHz ($\mu = 0, 1, \ldots, 4$) with up to 3300 subcarriers and proportionally reduced CP lengths. In FR1, subcarrier spacings of 15 kHz and 30 kHz are suitable, while in FR2, larger subcarrier spacings of 60 kHz and above are needed to handle, for example, the phase noise at millimeter wave frequencies.

With the use of CP-OFDM, NR naturally adopts OFDMA as its primary multiple access scheme where gNB assigns orthogonal radio resources to the served UEs. The controversy was centered around the roles of non-orthogonal multiple access (NOMA) in NR. NOMA allows multiple UEs to share the same radio resource so as to improve the system capacity.

The theoretical basis of NOMA is network information theory that models downlink and uplink NOMA as Gaussian broadcast channel and multiple access channel, respectively. It has long been known that the capacity regions of Gaussian broadcast channel and multiple access channel are larger than their orthogonal counterparts [7].

The potential gains of NOMA over OFDMA and the desire of achieving massive 5G connectivity have led to much research work and enthusiasm on NOMA. In the past several years, we have witnessed abundant NOMA proposals such as resource spread multiple access (RSMA), multi-user shared access (MUSA), sparse code multiple access (SCMA), and interleave division multiple access (IDMA). On a high level, different UEs are multiplexed on the same radio resource by using different power coefficients, codewords, sequences, and interleavers in these NOMA schemes.

3GPP has also carried out a comprehensive study on NOMA for NR. NOMA's theoretical promise did not translate into gains in more realistic scenarios, as detailed in TR 38.812 [8]. Hence, the study did not conclude the benefits of NOMA for

5G. Accordingly, 3GPP is only conducting minimal normative work on two-step random access as part of Release-16 NR specifications as a follow-up of the study.

## LTE-NR INTERWORKING AND COEXISTENCE

**Myth 3:** *With NR, LTE is not needed.*

The excitement about NR could easily lead to the illusion that LTE is obsolete. The fact is that LTE will remain the dominant mobile access technology in the next several years while NR is being rolled out. The initial NR deployments are taking place in areas with high traffic demand, and ubiquitous coverage will be provided by both LTE and NR in the next few years. Even after nationwide NR coverage is built, LTE will stay for a long time to support legacy devices. So, LTE and NR will coexist in the foreseeable future.

LTE-NR integration is an important design consideration of NR and has impacted many NR design aspects. Tight LTE-NR interworking is possible by LTE-NR dual connectivity starting with the first release of NR specifications. The LTE-NR dual connectivity is formally known as Evolved universal terrestrial radio access – New radio Dual Connectivity (EN-DC). In EN-DC, LTE evolved NodeB (eNB) is the master node, NR gNB is the secondary node, and a UE can have simultaneous connectivity to the eNB and the gNB. The eNB has a direct interface with the existing evolved packet core (EPC) network in both user plane (carrying user data) and control plane (carrying signaling traffic). The gNB however only has a direct interface with the EPC in the user plane. Thus, in this scenario NR provides additional user plane capacity but relies on LTE for connection management.

EN-DC is commonly known as NR non-standalone (NSA) mode. With EN-DC setup, NR operating at millimeter wave frequencies may be deployed as a small-cell layer overlaid with an LTE macro-cell layer operating in sub-6 GHz spectrum. EN-DC can also facilitate co-site deployment of NR with LTE to avoid the cost deploying new sites. Figure 2 gives an illustration of the EN-DC deployment option as well as the standalone NR deployment option.

NR has been designed to enable dynamic spectrum sharing with LTE when they are deployed in the same spectrum. Here we highlight a few key design aspects that make NR-LTE dynamic spectrum sharing possible.

- NR supports LTE-compatible numerology with 15 kHz subcarrier spacing, allowing for NR and LTE operations on a common time-frequency resource grid.
- The flexible NR design offers the possibility of placing the NR downlink control channels, initial access related reference signals, and data channels to minimize collisions with LTE reference signals.
- NR introduces reserved resources, which are semi-statically configured time-frequency resources for NR PDSCH to rate match around. There are three types of reserved resources: resources corresponding to LTE cell specific reference signals, resources corresponding to NR control resource sets, and resources whose positions on the time-frequency resource grid are indicated reserved by bitmaps. Using reserved resources is also an important tool for forward compatibility to allow for introducing new technologies and applications in the future.

The above NR features together with implementation based dynamic scheduling give the possibility of dynamically sharing spectrum between NR and LTE carriers based on traffic demand. They facilitate long-term LTE-NR coexistence and smooth migration of LTE to NR.

## LOW LATENCY SUPPORT

**Myth 4:** *5G delivers 1 ms end-to-end latency everywhere.*

Low latency support is a key 5G requirement. ITU-R defines minimum requirements on control plane latency (transition time from an idle state to the start of continuous data transfer) and user plane latency (one-way latency in the radio network) for eMBB and URLLC usage scenarios [9]. The control plane latency requirement is 20 ms for both eMBB and URLLC, while the user plane latency requirements are 4 ms for eMBB and 1 ms for URLLC. The often-quoted latency requirement for 5G is 1 ms over-the-air latency, which is sometimes misinterpreted as that 5G would deliver 1 ms end-to-end latency everywhere. The myth is certainly not true, for example, a radio signal traveling at the speed of light would have to spend ~13.8 ms purely on propagation from New York to San Francisco (the air distance of the two cities is about 4100 km).

Leaving aside the myth, low latency support is an important design target for NR. Fast scheduling is key to low latency. The typical scheduling unit is a slot, consisting of 14 OFDM symbols. With the possibility of using higher numerology in NR, slot duration scales down, i.e., NR slot duration is $2^{-\mu}$ ms for the subcarrier spacing of $2^{\mu} \cdot 15$ kHz ($\mu = 0, 1, \ldots, 3$). When $\mu = 0$, the slot duration is 1 ms, which is the same as in LTE. The shorter slot duration offers the possibility of faster slot-based scheduling. Besides, NR control signaling and reference signals can be front-loaded at the beginning of the transmission. This helps reduce delay because UE can start to process and decode the received signal immediately without the need of buffering.

In addition to slot-based scheduling, NR introduces the so-called "mini-slot" scheduling for achieving lower latency. A mini-slot has a flexible start position in a regular slot and can be as short as 1 OFDM symbol. Thus, traffic with low-latency requirements can be more promptly scheduled without the need of waiting for the start of a slot boundary.

Fast scheduling needs the support of fast UE processing time. NR tightens UE processing time requirements, especially the PDSCH processing procedure time and the PUSCH processing procedure time. Release-15 NR supports two levels of UE processing capability: basic UE processing capability 1 and more advanced UE processing capability 2. Reducing the PDSCH processing procedure time allows for rapid hybrid automatic repeat request (HARQ) acknowledgement. Shortening the PUSCH processing procedure time leads to fast UE response to gNB scheduling of uplink traffic.

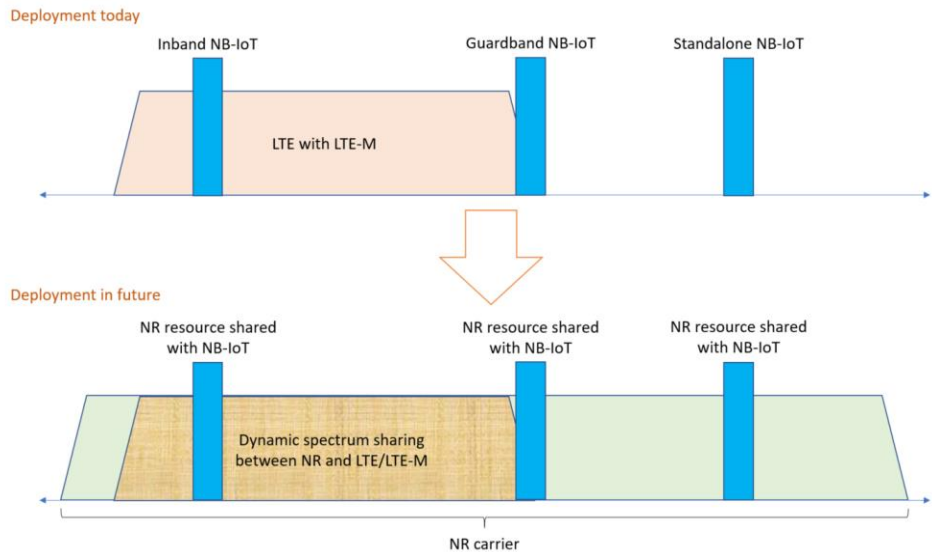

*Figure 3: An illustration of smooth migration of LTE/LTE-M/NB-IoT to NR deployment.*

NR higher-layer protocols have also been optimized for lower latency. For example, NR removes radio link control (RLC) concatenation that assembles multiple service data units (SDUs) into a single RLC protocol data unit (PDU). This enables the UE to preprocess RLC PDUs before receiving the uplink grant to meet the shortened PUSCH processing procedure time. NR also improves the medium access control (MAC) header structure, where the sub-header of a MAC SDU is placed right before the SDU. If all the MAC header information would to be placed at the beginning of a MAC PDU as in the case of LTE MAC, UE cannot start to assemble the MAC PDU until receiving the uplink scheduling grant.

## MASSIVE MACHINE TYPE COMMUNICATIONS

**Myth 5:** *NR is the only 5G access technology to serve massive machine type communications.*

As one of the three main usage scenarios envisioned for 5G, mMTC features a massive number of connected devices that are of low cost, low complexity, and long battery life. Example devices are sensors, actuators, and water/gas/parking meters. The volume of the traffic generated by each device is relatively low, and the applications are usually delay tolerant. But the devices may be in coverage challenging locations.

Since the NR interface is flexible and highly capable, it is tempting to think that NR is the only 5G access technology for addressing the mMTC usage scenario. But NB-IoT and LTE-M technologies are also an integral part of 5G.

3GPP specified NB-IoT and LTE-M in Release 13. Since then NB-IoT and LTE-M have been continuously evolved with new enhancements introduced in later releases [10] [11]. NB-IoT and LTE-M are optimized air interfaces tailored to LPWA communications for supporting the mMTC use case. For the mMTC use case, ITU requires that IMT-2020 should be capable of supporting a connection density of 1 million devices per square kilometer. 3GPP has shown that both NB-IoT and LTE-M can meet the IMT-2020 connection density requirement [12] and proposed both technologies to ITU-R as an integral part of 5G.

NR can meet the IMT-2020 connection density requirement as well to support mMTC usage scenario [12]. But the NR specifications of Releases 15 and 16 focus on eMBB and URLLC usage scenarios. Thus, NR has not yet been optimized for mMTC usage scenario and does not provide the extended coverage as delivered by NB-IoT and LTE-M. To fully address the diverse 5G mMTC use cases, it is of importance to combine NR deployment with NB-IoT and LTE-M deployments.

NR can well coexist with NB-IoT and LTE-M for reasons similar to how NR can coexist with LTE. The air interfaces are all based on OFDM and support 15 kHz numerology, making it possible to operate them on a common time-frequency resource grid. NB-IoT and LTE-M can be placed inside an NR carrier. The resources used for NB-IoT and LTE-M can be configured as reserved to NR UEs. This allows for smooth migration to NR while continuing to support the deployed NB-IoT and LTE-M devices that will remain in service for many years due to the long battery lives (e.g., 10 years) of these devices. Figure 3 gives an illustration on how the deployments of LTE/LTE-M/NB-IoT today can be smoothly migrated to NR deployments.

## NON-TERRESTRIAL COMMUNICATIONS

**Myth 6:** *NR is all about terrestrial mobile communications.*

The focus of 5G has been on terrestrial communications. NR, as the new air interface for 5G, has accordingly been designed with terrestrial communications in mind. Nonetheless, later releases of NR will continue the evolution, and enabling NR to support NTN is one direction under exploration in 3GPP [13]. Thus far, 3GPP has completed several studies on NTN use cases, requirements, channel models, deployment scenarios, and potential key impacts on NR.

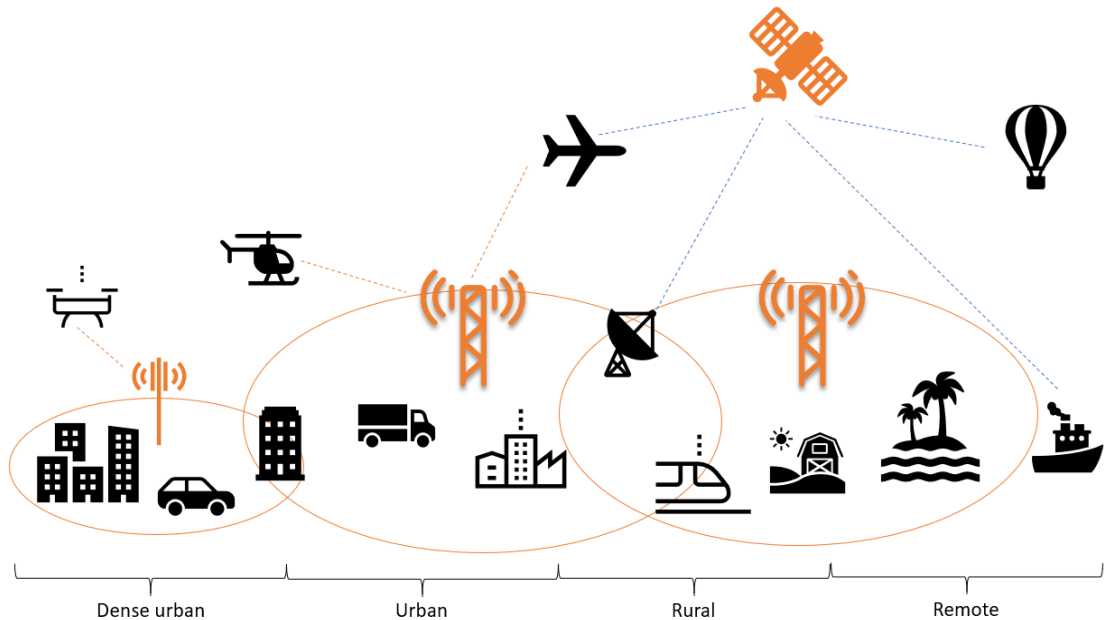

*Figure 4: An illustration of ubiquitous 5G connectivity by complementing terrestrial networks with non-terrestrial communications.*

To evolve NR to support NTN, there are three main challenges that shall be addressed: moving cells, long propagation delays, and large Doppler shifts.

As in any previous generation of mobile communications, the default assumption in NR design is that cells are stationary. However, this fundamental assumption is not true in NTN, particularly when the NTN has low-earth orbit (LEO) satellites as its spaceborne platforms. A LEO satellite is visible to a ground UE for only a few minutes. The spotbeam serving the UE may switch every few seconds if the spotbeams are moving with the satellite. Dealing with moving cells requires a rethinking of many of the NR design aspects. Take idle UE mobility management as an example. In order to page UE in idle mode, the network needs to know the UE position at least at a tracking area level. A tracking area is a cluster of gNBs or cells. To track UE in idle mode, the UE is provided with a list of tracking areas. The UE performs a tracking area update to notify the network if it moves and camps on a cell that does not belong to any tracking area in the provided list. If the tracking areas in NTN sweep over the ground with the satellites, a stationary UE would have to keep performing tracking area updates in idle mode, which should certainly be avoided. Similarly, connected mode mobility management also requires a rethinking to avoid the frequent handovers in NTN due to moving cells.

The propagation delays in terrestrial mobile systems are usually less than 1 ms. In contrast, the propagation delays in NTN are much longer, ranging from several milliseconds to hundreds of milliseconds depending on the altitudes of the spaceborne or airborne platforms in NTN. Dealing with such long propagation delays requires modifications of many timing aspects in NR from physical layer to higher layers.

Large Doppler shifts are another key issue resulted from the movements of the spaceborne or airborne platforms in NTN. For example, a LEO satellite at the height of 600 km results in a time-varying Doppler shift as large as 24 ppm (amounting to 48 kHz at 2 GHz carrier frequency). UEs located in different locations in a spotbeam experience different Doppler shifts. How downlink and uplink synchronizations should be performed in the presence of such large, fast varying Doppler shifts is a key design issue for NR which is based on OFDM and uses OFDMA as its multiple access scheme.

In addition to providing connectivity from space to ground, terrestrial networks can also provide connectivity to UAVs in the sky besides serving terrestrial UEs. Already today, LTE networks can support the initial deployment of low-altitude UAVs [14]. The more flexible, powerful NR air interface will deliver more efficient and effective connectivity for wide-scale UAV deployments.

NTN can complement the terrestrial networks to provide coverage, for example, to certain rural and remote areas where terrestrial coverage is not available. As illustrated in Figure 4, integrating NTN into terrestrial 5G networks is instrumental in providing ubiquitous 5G connectivity.

## BEYOND RADIO

**Myth 7:** *5G = NR.*

While previous generations of mobile communications systems were very much radio focused, 5G goes much beyond radio and thus is not just about NR. The other parts of 5G systems including core network, backhaul, and fronthaul are also being transformed to achieve much more efficient networks and enable new services [15].

3GPP has developed a new 5G core network (5GC) that has a service-based architecture. 5GC features end-to-end flexibility by separating the software functions from the core network hardware. Software defined networking/network functional virtualization, network slicing, and cloud-based architecture facilitate network softwarization.

Backhaul connects the radio access network to the core network. Fronthaul connects the remote radio units of a base station to the centralized radio controllers. The logical architecture of gNB is split into two parts called CU (Central Unit) and DU (Distributed Unit). The CU and DU are connected by a new interface called F1. 5G needs to deliver much higher data rates, lower latency, and greater capacity, compared to the previous generations of mobile communications systems. Accordingly, 5G backhaul and fronthaul need to be capable enough to accommodate the 5G technical requirements and should not become the bottleneck of the 5G systems.

## CONCLUSIONS

It is an exciting time in the wireless industry to witness the completion of the first releases of NR specifications and the rollout of 5G networks. The rightful excitement sometimes comes with confusion about what NR is (and is not). In this article, we have attempted to debunk seven of the more popular myths. We have summarized the key features of NR including spectrum flexibility, flexible waveform and multiple access, LTE-NR interworking and coexistence, and low latency support. We have pointed out that NB-IoT and LTE-M are an integral part of 5G and will coexist with NR. We have also highlighted that NR will continue evolution with NTN being an interesting direction. Finally, we have emphasized that not just the radio but the entire system of 5G is being transformed towards enabling new services and new ecosystems.

## BIOGRAPHY

**Xingqin Lin** is a Senior Researcher and Standardization Delegate at Ericsson, leading 4G/5G research and standardization in the areas of unmanned aerial vehicles (UAVs) and satellites. He is a key contributor to 5G NR, NB-IoT, and LTE standards. His pioneering work on cellular connected UAVs helped establish the 3GPP Rel-15 work on enhanced LTE support for aerial vehicles. He is a highly experienced telecom professional with expert knowledge in wireless communications and technology strategy. He is a frequent speaker, panelist, and technical contributor at conferences and workshops. He served as an editor of the *IEEE COMMUNICATIONS LETTERS* from 2015-2018. He holds a Ph.D. in electrical and computer engineering from The University of Texas at Austin, USA.